\documentclass[preprint,review,12pt]{elsarticle}

\newif\ifblind
\blindfalse 

\usepackage{amssymb}
\usepackage{amsmath}
\usepackage{xcolor}
\usepackage{soul}

\journal{Extreme Mechanics Letters}

\begin{document}

\begin{frontmatter}



\title{Multi-level mechanical modeling and computational design framework for weft knitted fabrics}


\ifblind
  \author{Anonymous Author(s)}
  \address{Affiliation withheld for double-anonymized review}
\else
    \author[label1]{Cosima du Pasquier\corref{cor1}}
    \ead{cosimad@stanford.edu}
    \affiliation[label1]{organization={Collaborative Haptics and Robotics in Medicine Lab (CHARM), Department of Mechanical Engineering, Stanford University},
                city={Stanford},
                postcode={94305},
                state={CA},
                country={USA}}
      
    \author[label1]{Sehui Jeong\corref{cor1}}
    \cortext[cor1]{CdP \& SJ contributed equally to this work.}
    \author[label3]{Pan Liu}
    
    \author[label2]{Susan Williams}
    \affiliation[label2]{organization={Self-Assembly Lab, Department of Architecture, Massachusetts Institute of Technology (MIT)},
                city={Cambridge},
                postcode={02139},
                state={MA},
                country={USA}}

    \author[label3]{Nour Mnejja}
    
    \author[label1]{Allison M. Okamura}
    
    \author[label2]{Skylar Tibbits}
    
    \author[label3]{Tian Chen}
    \ead{tianchen@uh.edu}
    \affiliation[label3]{organization={Architected Intelligent Matter Laboratory, Department of Mechanical and Aerospace Engineering, University of Houston},
                city={Houston},
                postcode={77204},
                state={TX},
                country={USA}}
\fi

\begin{abstract}
This work presents a multi-level modeling and design framework for weft knitted fabrics, beginning with a volumetric finite element analysis capturing their mechanical behavior from fundamental principles. Incorporating yarn-level data, it accurately predicts stress-strain responses, reducing the need for extensive physical testing. A simplified strain energy approach homogenizes the results into three key variables, enabling rapid, accurate predictions in minutes. After validation against experiments, our framework can simulate new knit fabrics without additional tests. In real-world scenarios, fabrics often feature variations in yarn materials or patterns. The framework extends to heterogeneous fabrics, showing that transitions between distinct regions can be captured using simple mechanical analogies: springs in series and parallel. This allows heterogeneous textiles to be treated as idealized patchworks of homogeneous pieces, preserving predictive accuracy. The method is demonstrated by designing and producing a compression sleeve with uniform pressure, illustrating how the framework supports development of knits tailored to specific assistance levels and anatomical features. By combining volumetric finite element analysis, simplified model through homogenization, and controlled material transitions, this approach provides a scalable, high-fidelity path toward next-generation weft knitted fabric design.
\end{abstract}

\end{frontmatter}



\section{Introduction}
\begin{figure}[t]
    \centering
     \includegraphics[width=\textwidth]{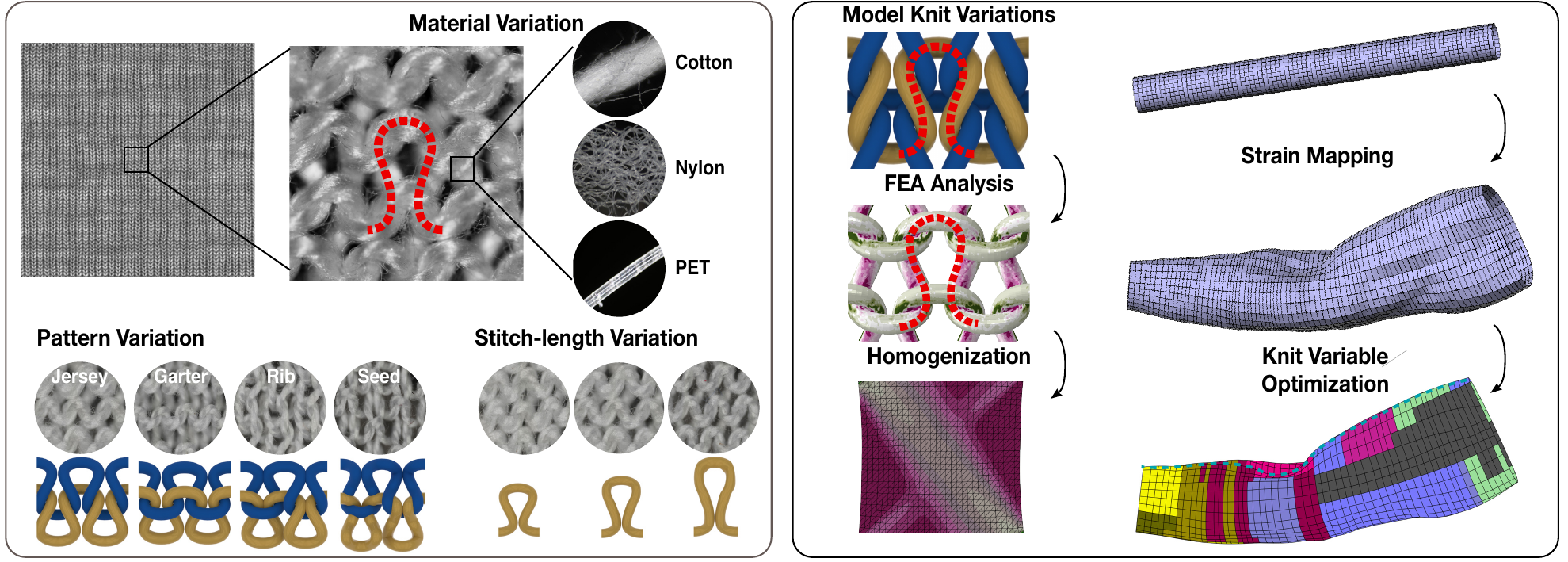}
      \caption{Principles of knitted fabrics. (Left) Knits are made of a hierarchical microstructure, starting from bundles of fibers; their response can be modulated using three variables: material of the fibers, pattern, and stitch-length. (Right) By geometrically reconstructing the knit patterns, we use the material properties of the yarns to build a volumetric finite-element analysis of a representative knit unit of 2 x 2 knots; this behavior can be generalized to simulate homogeneous knit samples and eventually to optimize the stress and strain in a garment - a compressive sleeve with even pressure distribution.}
     \label{fig:graphical_abstract}
\end{figure}
In the rapidly evolving landscape of wearable technology and soft robotics, the demand for materials that seamlessly integrate flexibility, durability, and functionality has never been greater. Knit fabrics, with their inherent stretchability and adaptability, have emerged as prime candidates to meet these requirements~\cite{sanchez20233d,Collier1999,Luo2022,duPasquier2024,singal2024programming}. Their unique and complex microstructure are the basis for their versatility in mechanical response, ultimately enabling weft knits to endure repeated large stretches without adhesives or
additives~\cite{warren2018}. While weft knitting itself has remained relatively unchanged, industrialization and digital fabrication have significantly increased the complexity, precision, and scalability of the method. 

Despite these technological advances, the full potential of knit fabrics remains limited by an incomplete understanding of their complex mechanics. Knits exhibit highly nonlinear and history-dependent behaviors due to intricate yarn interactions involving bending, stretching, frictional sliding, and jamming~\cite{postle2002structural,Bueno2008}. These complexities present significant modeling challenges, particularly in heterogeneous or multi-layered knit structures encountered in real-world applications.

Foundational efforts in modeling knit mechanics started with geometric methods introduced by Peirce~\cite{Peirce1947} and refined by Leaf~\cite{Leaf1955,Leaf1960}, focusing primarily on predicting fabric dimensions and stable loop configurations. Empirical contributions by Munden further established the crucial role of stitch length as a primary geometric parameter~\cite{Munden1959}. However, these early models lacked essential mechanical realism, neglecting yarn elasticity and friction. Later, Postle's energy-based approaches improved realism by incorporating yarn bending and tension energies, significantly enhancing predictive capabilities~\cite{postle2002structural}. Contemporary research, including recent studies by Poincloux et al., has further expanded these principles into nonlinear and dynamic domains, using sophisticated numerical simulations and advanced homogenization techniques to bridge microscopic yarn-level mechanics with macroscopic fabric behavior~\cite{Poincloux2018,weeger2018nonlinear,dinh2018prediction,Huang2024}. Building on prior modeling approaches, Singal et al. proposed a multiscale framework that connects yarn-scale mechanics with continuum-level constitutive models, providing predictive insight into the elasticity of knitted fabrics\cite{singal2024programming}.

Today, computational modeling methods are categorized into qualitative and quantitative approaches. Qualitative methods prioritize visual accuracy and fabric animation through techniques such as image feature extraction, graph-based mesh generation and software tools like CLO3D~\cite{Kaldor2010,singal2024programming,Sperl2020,Kaldor2008,Cirio2017,Huang2022}. In contrast, quantitative modeling emphasizes mechanical fidelity, incorporating yarn-scale elasticity, contact, and friction to predict stress–strain responses~\cite{Poincloux2018,McKee2017,Grandgeorge2021,Baek2021,abel2012two,Liu2019}. Recent advances in this domain have employed high-fidelity finite element models to resolve yarn bending, contact, and friction at the loop scale~\cite{Cherradi2022MechanicalValidation:,Liu2019}, with seminal contributions by Liu et al. establishing the role of material architecture and interfacial effects in determining mechanical response at the yarn level~\cite{Liu2017,Liu2018}. These approaches, together with representative unit cell methods~\cite{dinh2018prediction}, enable powerful predictions but are usually limited to uniform, homogeneous fabrics and require extensive calibration for each new configuration.

Building on these quantitative approaches, this work presents a comprehensive framework that bridges yarn mechanics and multi-scale textile simulation. We introduce a finite element-based volumetric model of knit stitches that incorporates anisotropic yarn behaviors and contact interactions. Unlike previous yarn-level models, our framework is experimentally validated across a systematic set of swatches with varying stitch lengths, patterns, and yarn materials, enabling accurate generalization without per-fabric recalibration. The results inform a simplified, reduced-order strain energy model compatible with commercial FEA tools, allowing efficient simulation of fabric-level responses. Critically, this framework extends to heterogeneous textiles, incorporating simple yet effective mechanical analogies to capture transitions across regions of different yarns or patterns. As a demonstration, we apply this framework to optimize a heterogeneous knitted compression sleeve that adapts to arm geometry while delivering spatially uniform pressure (see Fig.~\ref{fig:graphical_abstract}). By integrating physically grounded yarn-level modeling with scalable computational tools, this work addresses key limitations in current knit modeling efforts by improving predictive accuracy across material and pattern variations and establishing a practical foundation for engineering the next generation of functional knitted textiles.

\section{Mechanics of Yarns.} 
While prior studies have modeled knits using homogenized spring networks or detailed yarn-level finite element models, many assume simplified yarn behavior (e.g., linear, isotropic) or require extensive calibration for each fabric. In contrast, our approach begins by directly characterizing the anisotropic mechanics of industrial yarns, enabling a physics-based foundation for scalable, predictive modeling of knit fabrics with varying structure and material.

We select three industrial yarns spanning a wide mechanical spectrum: a 20-count, 2-ply cotton yarn (Supreme Corporation, 20/2), a multifilament, textured nylon yarn (Hickory Throwing, 70 denier, 24 filaments, 2-ply), and a PET monofilament (The Thread Exchange, 0.178 mm diameter). Yarns are generally either plied bundles of microfibers or uniformly extruded monofilaments. Plied yarns include both natural and synthetic fibers: natural fibers (e.g., cotton) are short-staple, while synthetic fibers are typically continuous filaments. To capture contrasting mechanics, we selected cotton (strain capacity $<10\%$) and nylon (elongation $>120\%$). As a representative monofilament, we also included extruded PET~\cite{Bueno2008}.

\begin{figure}[htbp]
\centering
 \includegraphics[width=0.75\textwidth]{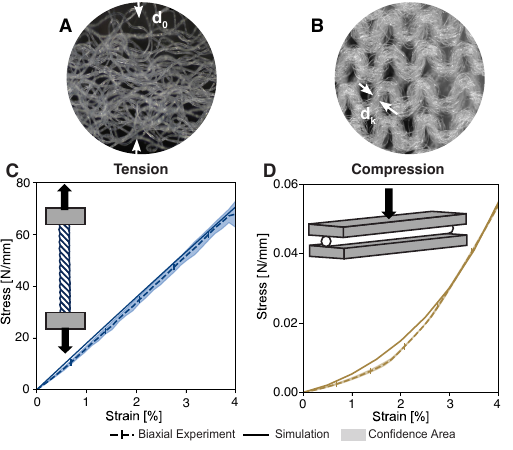}
 \caption{Yarn testing methodology; microscopy images of nylon A. yarn and B. knit fabric sample; illustration of test procedure and results for cotton yarn testing in C. tension and D. compression; the confidence area represents $\pm$ one standard deviation from the mean, computed from $n=3$ samples.}
 \label{fig:method_overview_yarn}
\end{figure}

Recognizing that yarns may exhibit strong anisotropy from internal architecture (e.g., twist, voids, filament alignment), we experimentally quantify both axial and transverse mechanical responses ~\cite{singal2024programming,el2023analysis}. Experimental details, including specimen description, preconditioning, and strain rate, are provided in SI Section~1. Results show that cotton and nylon behave as transversely isotropic materials, with transverse stiffness orders of magnitude lower than their longitudinal stiffness. This pronounced anisotropy arises from the reconfiguration of loosely bound fibers under compression (Fig.~\ref{fig:method_overview_yarn}A, B, and D). By contrast, PET behaves as an isotropic monofilament.

To capture the effects of tension introduced during knitting, we distinguish the free-standing yarn diameter ($d_0$) from the post-knit diameter ($d_k$), with the difference indicating pre-stress (see Fig~\ref{fig:method_overview_yarn}A and B). For compression tests, yarns were pretensioned to match this in-fabric state, ensuring accurate mechanical behavior in subsequent simulations.

Using this experimental data, shown in Fig~\ref{fig:method_overview_yarn}C and D, we homogenize the yarn behavior and represent each stitch as a solid curved rod within a volumetric FEA model. This approach bridges the micro-scale anisotropic mechanics of the yarns with the meso-scale knit structure, enabling accurate prediction of mechanical response across fabric variations.

\section{Mechanics of homogeneous knit fabric}
Industrial knit fabric typically consists of various homogeneous knit units, each defined by specific knit parameters to achieve target mechanical behaviors, such as stiffness and stretchability. We characterize homogeneous knit fabrics and establish an experimental protocol suitable for numerical modeling. We conduct a volumetric FEA and validate it against experimental data, then propose a reduced-order model that efficiently represents the anisotropic and nonlinear behavior of knits. To support this, we conducted preliminary tests to identify key factors influencing mechanical response, including cyclic loading, post-processing (washing and drying), inter-yarn friction, viscoelasticity (strain-rate dependence), sweat exposure, and stress concentrations from geometry. Details on the fabrication of samples are provided in SI Section 2 and the Appendix.

\subsection{Preliminary Testing.}
In preliminary testing (SI Section 3), we systematically examined key factors affecting knit mechanics, primarily using jersey cotton specimens with stitch length 11 (our benchmark for this work) due to their widespread use and consistent behavior. We observed significant initial softening and hysteresis under cyclic loading, consistent with Mullins-like effects, and rapid stabilization during long-term fatigue tests. Relaxation (washing and drying) partially restored fabric stiffness. We selectively included nylon and PET knits in friction and viscoelasticity tests to investigate effects specific to their distinct yarn structures—extruded textured fibers and monofilament versus plied yarn. Friction tests showed minimal impact of lubrication, except initially for nylon. Viscoelastic effects were minor overall, notable only in PET monofilament samples at higher strains. Artificial sweat exposure minimally affected fabric behavior within typical clothing-use strains ($<$15\%). Finally, we concluded from these preliminary tests to use cruciform specimens to mitigate localized damage from stress concentrations at higher strains ($>$10\%).

\subsection{Knitting Variables.}
Two fundamental manufacturing parameters~\cite{Bueno2008}—stitch length and knitting pattern—are examined in addition to yarn material. Stitch length (SL), controlled by the tension during knitting, influences the fabric’s density and mechanical behavior. Three SLs are considered (loose, tight, very tight), corresponding to machine settings of 10, 11, and 12. Each stitch has an estimated arc length of 3.9~mm, 4.7~mm, or 5.9~mm for a jersey pattern, based on the representative parameterized geometry described in SI Section~7.A.1. Geometric parameters were measured from $16\times$ microscopy images of 12 stitches, the curve was fit to these measurements, and the resulting arc length estimates were averaged. Actual values may vary with the specific knitting machine and pattern.
Knitting patterns result from specific arrangements of “knit” and “purl” stitches. A Jersey pattern uses identical stitches in each row, Garter alternates rows, Rib alternates columns, and Seed alternates both (see Graphical Abstract). Column alternations (in the course direction) involve horizontal runs of yarn, while row alternations (in the wale direction) align vertically.
Sample preparation and testing procedures are in the Appendix and our biaxial testing stage in described in SI Section 4. 

\subsection{FEA modeling}
\begin{figure}[htbp]
\centering
\includegraphics[width=0.65\textwidth]{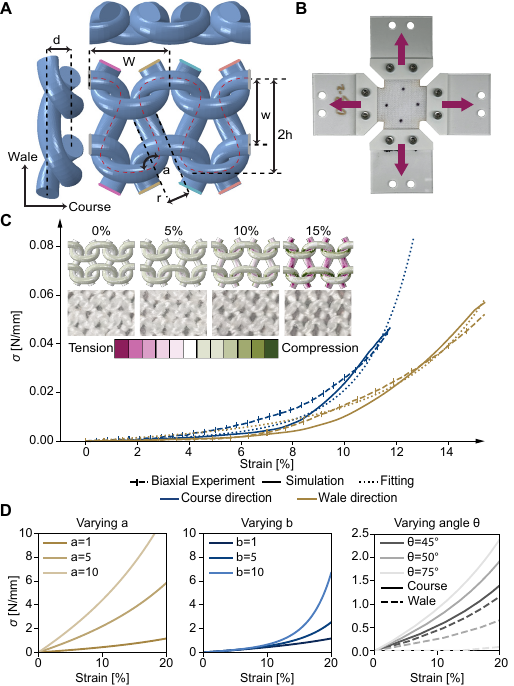}
\caption{Fabric swatch numerical modeling and experimental methodology; A. Finite Element (FE) model showing the geometry of a 2$\times$2 knit stitches (2 stitches in the course and the wale direction); parameters for the centerline, and color-coded boundaries that are tied to enforce periodicity; B. Biaxial experimental stage for biaxial testing of the cruciform fabric swatches; global stress is measured using load cells and local strain is measured using four markers and digital image correlation; C. Normalized Force-Displacement plot of a benchmark knit sample in both X (course) and Y (wale) directions from the FE simulation, reduced-order model, and experiment ($n=3$). Qualitative comparisons between FE and experiments are shown inset at different normalized displacements. In the FE, the color represents the maximum absolute principal stress, ranging from 6 MPa in tension to –4 MPa in compression, where white indicates no stress. D. Effect of the fitting parameters a, b, and $\theta$ on the stress-stretch response.}
\label{fig:fabric_testing_methodology}
\end{figure}

Our FE model captures the mechanical behavior of knits with different stitch lengths, pattern and yarn material. The knit geometry and topology is defined using explicit parametric equation of the yarn centerline~\cite{crane2023simple}. Instead of modeling individual fibers, the mechanical behavior of the yarns is homogenized based on the aforementioned testing procedures: PET as an isotropic material; cotton and nylon as transversely isotropic  materials~\cite{Poincloux2018,singal2024programming}. Detailed model definition including centerline geometry, material model, and boundary conditions are described in SI Section 7.A.
\begin{itemize}
\item \textit{Geometry and Meshing:}
A centerline describing each stitch is modified from a trigonometric curve for Jersey \cite{crane2023simple} and adapted to other patterns. A solid geometry is extruded along this centerline and meshed with quadratic tetrahedral elements (C3D10) using a seed size of $1/3$ of the initial radius. Microscopic images guide the specific geometry parameters.

\item \textit{Representative Unit Cell:}  
Each pattern has a repeating unit to exploit periodicity (Fig.~\ref{fig:fabric_testing_methodology}A). A 2$\times$2 arrangement of stitches sufficiently captures contact, friction, sliding, and potential cross-sectional changes. Periodic boundary conditions (matching pairs of nodes and edges) replicate the repeated nature of the knit.

\item \textit{Pre-stress Step:}  
To prevent any initial interference for mesh validity, we initialize the yarns with a smaller diameter. We then apply radial thermal expansion (a standard FE technique \cite{Smith2014}) to gradually increase the diameter until it reaches the actual yarn diameter without changing the length. This approach replicates the diameter variations induced by contact and tension during knitting.

\item \textit{Biaxial Testing:}  
Numerical load steps replicate the equibiaxial stretching applied to cruciform samples in experiments (Fig.~\ref{fig:fabric_testing_methodology}B) using a custom biaxial stage, following a similar test procedure as Connolly et al.~\cite{Connolly2019}. Details of the experiments, including the setup specifications, specimen preparation, and loading rate, are provided in SI Section~4. Reaction forces in the course (X) and wale (Y) directions are extracted and compared to experimental data. For brevity, we report results in the course direction only; wale direction results are provided in SI Section 6.

\end{itemize}
\subsubsection{Numerical validation of the FEA.}
Stress-strain curves from FE simulation are compared with experimental results. Normalized root mean squared error (NRMSE), normalized by the maximum measured stress, is generally below 5\%. One exception occurs for stitch lengths of 10 (9.43\% and 12.4\% simulation error in course and wale respectively), arising from tight contact interference and model compromises for fitting both fabric directions. The full results are reported in SI Section 6A.

Given that the simulations match experiments within acceptable bounds, the volumetric FEA framework can be used to predict mechanical behavior for new or modified knit designs without further physical testing. This result significantly accelerates and simplifies the development process, by replacing fabricating and experimentally testing every design iteration.

\subsection{Strain Energy Model for Homogenized Knit Behavior}
Although the volumetric FEA offers detailed insight, it is computationally expensive (of the order of tens of hours). For the computational design and optimization of knit fabrics, a simplified strain energy model is introduced that captures the essential stiffness and anisotropy of the fabric with only three parameters: $a$, $b$, and $\theta$. These parameters directly map to physically meaningful properties, with  relating to initial stiffness,  overall stiffness growth, and  direction of anisotropy. The formulation is based on the anisotropic part of the Holzapfel-Gasser-Ogden (HGO) model \cite{holzapfel2000new}, already implemented in commercial FE tools such as ABAQUS. The derivation is detailed in SI Section 7.B.

\begin{equation} \psi=\frac{a}{b}\left[\exp\left(b\left(I_{4}-1\right)^2\right)-1\right],
\label{eq:SEmodel} 
\end{equation}

\noindent is the strain-energy, where $I_{4}$ is the fourth invariant, derived from the deformation tensor $\boldsymbol{F}$ and an in-plane anisotropy direction $\boldsymbol{N}=[\text{cos}\theta, \text{sin}\theta, 0]^T$. 

\begin{equation} I_{4} = \left[\boldsymbol{F}^{T}\cdot\boldsymbol{F}\right]:\boldsymbol{N} = {\lambda_{1}}^2\cos^{2}\theta+{\lambda_{2}}^2\sin^{2}\theta , \end{equation}

\noindent where in-plane directions correspond to the course ($\lambda_{1}$) and wale ($\lambda_{2}$) stretches. The stress (first Piola-Kirchhoff) then follows:

\begin{multline}
    \boldsymbol{P}=\frac{\partial\psi}{\partial \boldsymbol{F}}=\frac{\partial\psi}{\partial  I_4}\frac{\partial I_4}{\partial \boldsymbol{F}} \\
    =2a(I_4 - 1)\exp\left(b\left(I_{4}-1\right)^2\right)\left[2\lambda_1\cos^2\theta,2\lambda_2\sin^2\theta\right]^T.
\label{eq:PiolaStress}
\end{multline}

In the low-stress regime with $I_{4} \approx$ 1, this strain energy can be approximated by $\psi=a\left(I_{4}-1\right)^2$. As illustrated in Eqs.~\ref{eq:SEmodel} and \ref{eq:PiolaStress}, both the strain energy and stress expressions include terms from the course and wale directions ($\lambda_1, \lambda_2$). This indicates that deformations in the course and wale directions are coupled.

We compare the behavior of fabrics depending on stitch length, pattern, and material using $a$, $b$, and $\theta$. The parameters are fitted to the stress-strain curves obtained experimentally and through simulations. We use the Mesh Adaptive Direct Search (MADS) algorithm to find the optimal parameters by minimizing the squared-difference of the simplified model and the stress-strain response~\cite{nomad4paper}:
\begin{equation}
    \underset{\boldsymbol{x}}{\mathrm{argmin}}\underset{i\leq n}{\sum}\|\boldsymbol{S}_{i}-\boldsymbol{P}_{i}\left(\boldsymbol{x}\right)\|^2 
\label{eq:optimization} \end{equation}
where $n$ is the sample size, $\boldsymbol{x} = [a, b, \theta]^t$, $\boldsymbol{S}_{i}$ represents the measured nominal stress for the $i$th sample, and $\boldsymbol{P}_{i}$ denotes the resulting nominal stress from the strain model with the given parameter $\boldsymbol{x}$.

This reduced-order model reproduces the essential knit behavior with minimal computational effort. Under typical loading ranges, parameter fitting requires only a few minutes. Figure~\ref{fig:fabric_variation_testing} illustrates a benchmark comparison among experimental data, FEA predictions, and the simplified strain energy model.

\subsubsection{Results from the Systematic Testing of Knitting Variables} 
\begin{figure}[htbp]
\centering
\includegraphics[width=\textwidth]{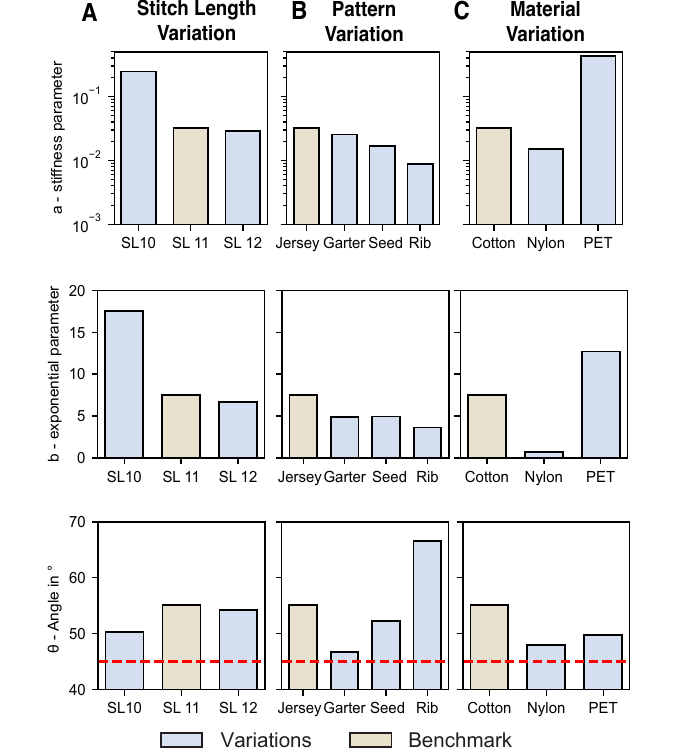}
\caption{Fitting parameters $a$, $b$ and $\theta$ of the strain energy model for three types of variations. The red line at $\theta = 45^\circ$ indicates identical stress in course and wale directions under equibiaxial loading.}
\label{fig:variations_strain_model}
\end{figure}

After preliminary loading stabilizes each specimen, two equibiaxial experiments on $n=3$ samples are conducted for each variation (stitch length, pattern, and yarn material). Figure~\ref{fig:fabric_variation_testing} shows representative stress-strain curves in the course direction, accompanied by simulation and simplified model fits (results for the wale direction appear in SI Section 6). Figure~\ref{fig:variations_strain_model} displays how the fitting parameters ($a$, $b$, $\theta$) change with each variable:

\begin{figure}[htbp]
\centering
\includegraphics[width=\textwidth]{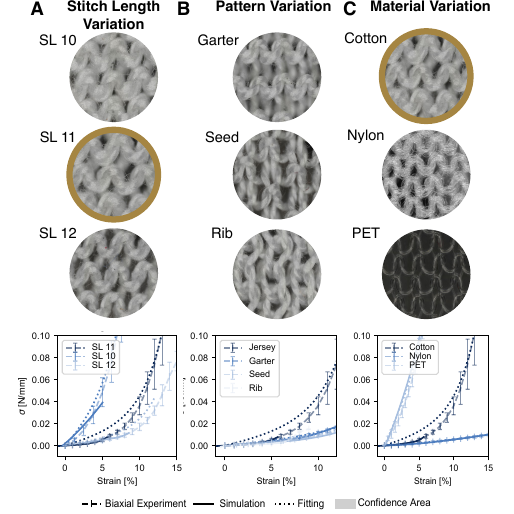}
\caption{Varying microstructure and mechanical response of three types of knit variation in the course direction: A. stitch length, B. pattern, and C. yarn material; the benchmark SL11, Jersey, Cotton is represented twice with a yellow outline; the normalized force-displacement curves show the experimental, simulated, and fitted responses for three variations in each category; the confidence area represents $\pm$ one standard deviation from the mean, calculated from $n=3$ samples.}
\label{fig:fabric_variation_testing}
\end{figure}

\begin{itemize}
\item \textit{Stitch Length:} Primarily influences overall stiffness ($b$), with shorter stitches increasing fabric stiffness (Fig.~\ref{fig:fabric_variation_testing} and Fig.~\ref{fig:variations_strain_model} A).
\item \textit{Pattern:} Greatly affects anisotropy ($\theta$). Patterns like Rib or Seed alter the symmetry and hence the relative stiffness in the course vs. wale directions (Fig.~\ref{fig:fabric_variation_testing} and Fig.~\ref{fig:variations_strain_model} B).
\item \textit{Material:} Impacts both initial ($a$) and overall stiffness ($b$), spanning more than an order of magnitude, but does not strongly affect anisotropy (Fig.~\ref{fig:fabric_variation_testing} and Fig.~\ref{fig:variations_strain_model} C).
\end{itemize}

Together, these findings offer practical guidelines for designing homogeneous knit fabrics: yarn material sets the global stiffness range, stitch length fine-tunes stiffness, and pattern controls anisotropy.

\subsubsection{Numerical validation of the strain energy model.} The strain energy model was validated against experimental data using the same NRMSE metric as for the FEA. The normalized error was below 5\% for all cases except stitch lengths 10 and 12. The higher error of SL 10 reflects the tight contact, same as in the FEA, while the deviation of SL 12 is attributed to initial slack from longer stitches producing a flat, linear response in the low strain regime. The correspondence demonstrates that the strain energy model successfully replicates FEA results with a small fraction of the computational cost.

Having established both a detailed FEA and a rapid strain energy model for homogeneous knits, the next step is to examine transitions between different yarns or patterns. 

\section{Mechanics of heterogeneous knit fabrics}

\begin{figure}[htbp]
\centering
\includegraphics[width=\textwidth]{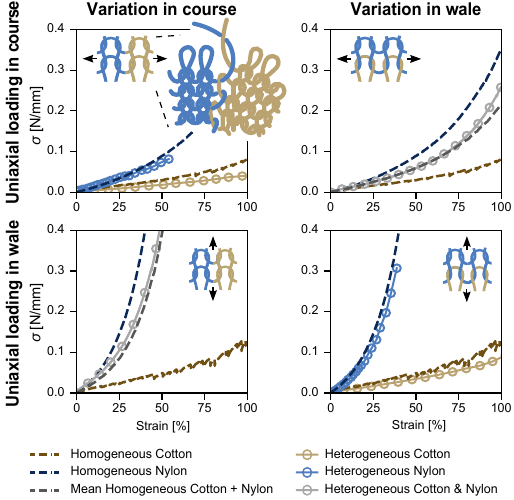}
\caption{Uniaxial testing results of homogeneous vs. heterogeneous swatches of cotton and nylon; material variation occurs along the course (right column) and wale (left column) direction; course transitions are done by intarsia, which is illustrated in the top left graph; testing occurs \emph{in parallel} (yarn interface parallel to loading axis) and \emph{in series} (yarn interface perpendicular to loading axis).}
\label{fig:heterogeneous-effect} 
\end{figure}

Building on the validated methodologies for homogeneous knits, we examine how transitioning knit parameters within the fabric affects overall mechanical behavior. Industrial knitting machines commonly introduce regional variations in stitch length, pattern, or yarn material to tailor a single textile to local functionality needs. Such \emph{heterogeneous} fabrics can, for instance, provide extra stiffness in high-load areas or enhanced stretch in regions requiring greater conformability.

\subsection{Material Transitions.}  
This section focuses on material transitions, which had the most pronounced impact on homogeneous swatch properties (pattern transitions appear in SI Section 6). Figure~\ref{fig:heterogeneous-effect} illustrates two types of transitions: along the course (right column) and along the wale (left column). A change in material in the course direction (i.e., from one column of stitches to the next) is achieved via an \emph{intarsia} technique, whereas transitions in the wale direction occur more seamlessly by knitting a different yarn across subsequent rows.

To isolate the effect of transitioning from one mechanical regime to another, square knit samples ($l,w=50\:\mathrm{mm}$) are prepared with one half of the swatch made from cotton and the other from nylon. The uniaxial testing procedure, including strain rates and gauge length, is described in SI Section~5. Each heterogeneous sample is uniaxially stretched to 100\% strain in both the course and wale directions (four tests total). Two testing modes are used:

\begin{itemize}
    \item \emph{In parallel}: The yarn interface aligns with the loading axis, so both materials share the load simultaneously.  
    \item \emph{In series}: The yarn interface is perpendicular to the loading axis, so the stretch transitions from one material to the other. Digital image correlation (DIC) tracks the displacement of the boundary between the two halves.
\end{itemize}

Homogeneous cotton and nylon swatches of similar dimensions are also tested under the same conditions for comparison.

\subsection{Results}
Figure~\ref{fig:heterogeneous-effect} shows that in the \emph{in parallel} configuration (top row), the stress-strain response of the heterogeneous sample closely matches the weighted average of the two corresponding homogeneous swatches. In other words, changing from cotton to nylon within the same fabric does not introduce significant deviations from a simple parallel-spring analogy. In the \emph{in series} tests (bottom row), the strain experienced by each material region also remains consistent with individual homogeneous behaviors, suggesting that the interface itself does not substantially affect the overall load distribution.

These findings confirm that material transitions in knit fabrics can be modeled accurately by treating the textile as a patchwork of homogeneous regions joined by simple mechanical constraints (e.g., springs in parallel or series). Thus, the strain energy models and FEA approaches developed for homogeneous fabrics extend to heterogeneous textiles, without major modifications to capture boundary effects. This greatly streamlines the design process for advanced, multifunctional garments, for which regional variations in pattern or yarn can be incorporated without substantially complicating mechanical predictions, as demonstrated in the next section.

One limitation of our model is that it only captures classic knit transitions, such as row transitions and intarsia (column) transitions, as shown in Fig.~\ref{fig:heterogeneous-effect}. The model does not describe uneven, slanted, or multi-layer transitions that introduce local structural variations. Future work will expand the framework to include these complex transition types.


\begin{figure}[htbp]
\centering
\includegraphics[width=0.8\textwidth]{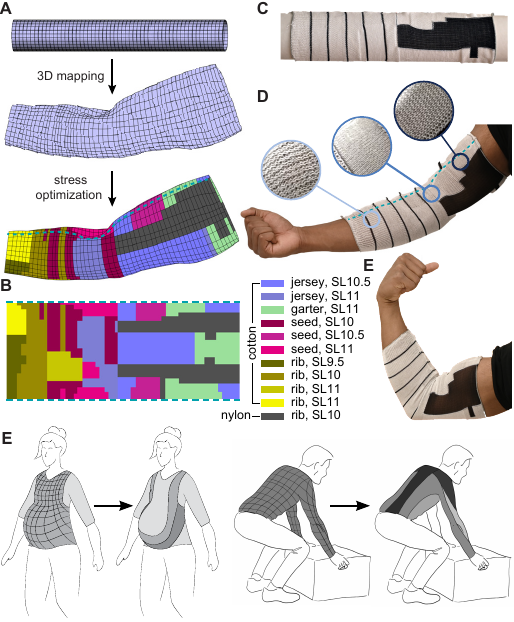}
\caption{Design of a tubular sleeve that results in constant stress magnitude when worn. A. A tubular geometry when knit with the wrist diameter as a reference is mapped to accommodate a muscular arm, after which knit variables are optimized to achieve uniform stress distribution; B. The resulting knit pattern with different regions assigned different stitch length, material and pattern; C. correspondence between the digital design and the physical specimen when stretched uniformly on a cylinder; D. The sleeve exhibiting variable stretch when worn on the arm, and the knit microstructure that enable this; E. the sleeve further accommodates muscle flexing when bend; the teal threaded line in A, B, and C indicates the seam position; F. Other potential applications of this approach in designing support garments for pregnancy or carrying heavy loads. }
\label{fig:fig-7_sleeve} 
\end{figure}

\section{Design of a uniform stress sleeve}
For skin-tight garments such as bodysuits, compression sleeves, and leggings, some stretch is present throughout the fabric to reduce wrinkling. To account for the varying stretch across different parts of the human body, \textit{e.g.}, different circumferences between the bicep and the wrist, additional panels are often sewn in. These extra panels reduce fabric damage from excess stretch and alleviate discomfort associated with large compressive stress to the body. (As shown earlier, generally, stress increases exponentially with stretch.) However, achieving uniform compression with paneling is challenging and produces fabric waste. The additional seams can also cause skin irritation and introduce weaknesses in the garments~\cite{Xiong2018}. Building on the results above, we demonstrate a single piece of simply shaped heterogeneous fabric that can accommodate spatially varying stretches while maintaining the uniform stress distribution throughout. Related work in computational design of knitted structures, such as Liu et al.’s “Knitting 4D Garments”~\cite{Liu2021}, has demonstrated how elasticity control enables garments to adapt dynamically to body motion. Our approach complements these efforts by focusing on predictive mechanical modeling and optimization to achieve uniform, passive compression without post-actuation.

As an example, we reconstruct the arm of a muscular volunteer in three dimensions. By generating a quadrilateral mesh and mapping it onto a regular cylinder with a radius matching that of the wrist, we derive the strain experienced at each material point on the arm (Fig.~\ref{fig:fig-7_sleeve}A). We then identify knit parameter combinations that produce consistent stress under these varying strains (Fig.~\ref{fig:fig-7_sleeve}B). We then conduct experimental measurements using a thin-film flexible force sensor (Tekscan, FlexiForce A201 with 1 pound capacity) at multiple points along the arm, before and after a 45-minute arm-focused weight-lifting exercise (see SI Section 9). The results demonstrate that our optimized knit sleeve maintains uniform compression under varying physiological conditions, confirming the accuracy of our computational optimization. This quantitative validation underscores the effectiveness of our single-piece heterogeneous knit fabric in delivering targeted, comfortable, and uniform compression.

\section{Conclusion}
We introduce a multi-level modeling strategy for industrial knits that combines a high-fidelity FEA framework with a homogenized surrogate model informed by experimental mechanics. This approach captures true mechanical performance and extends to heterogeneous knits, so that spatially varying properties can be achieved. By preserving predictive accuracy while reducing computational cost, the model makes optimization and large-scale design exploration of knits feasible. Our approach bridges fundamental mechanics and computational design, opening new frontiers to programmable textiles for wearables, medical devices, and soft robotics.

\appendix
\ifblind
\else
  \section*{Acknowledgments}
    We thank Lavender Tessmer for her guidance and early collaboration in industrial knitting. We thank Ushanth Balasuriya for volunteering his arm. Some of the computing for this project was performed on the Stanford University Sherlock cluster. We would like to thank Stanford University and the Stanford Research Computing Center for providing computational resources and support that contributed to these research results. We also thank Conor O'Brien at Otherlab for his advice and support in knitting the sleeve. 
  \section*{Funding}
\fi

\section{Manufacturing of knit specimens}
\label{app1}

We use a Stoll CMS 330 industrial knitting machine to knit the fabric swatches. The patterns are manually programmed except for the sleeve, which is generated algorithmically. Detailed information is given in SI Section 1.

\section{Yarn experimental testing}
\label{app2}
The yarns are washed and dried prior to mechanical loading. In tension, the yarns are loaded from the relaxed state (50 mm) until failure at a rate of 0.1 mm/s. The diameter of the yarns are measured throughout. In compression, the yarns are pre-tensioned until their diameter match those of the yarns in knit fabrics. They are then compressed at a rate of 0.01 mm/s up to a threshold of 50~N. A universal testing machine (Instron 68SC) is used in all tests. Additional details on the procedure are given in SI Section 2.

\section{Fabric swatch experimental testing}
\label{app3}
The homogeneous specimens are laser cut to a cruciform shape such the loads are applied to a 30 $\times$ 30 mm gauge area, except for the PET samples which are cut to a square shape due to unravelling at the corners of the cruciform.

All specimens are washed and dried prior to mechanical testing. Optical microscopy is used to obtain the geometrical parameters for FEA. All homogeneous specimens are tested using a custom biaxial testing stage (Fig.~\ref{fig:fabric_testing_methodology}B). 
The specimens are clamped on all four edges. Equibiaxial strain is applied up to 30\%. The actual strain in the gauge area is captured with an overhead camera. Three specimens are loaded cyclically three times each for each parameter variation.

Uniaxial testing is conducted to measure the influence of the different transitions (heterogeneous samples) using specimens of 50 $\times$ 50 mm.
Further details on the testing procedures can be found in SI Sections 3, 4, and 5. Both are based on previous testing procedures described by Connolly et al.~\cite{Connolly2019}.

\section{Numerical modeling of knits}
\label{app4}
ABAQUS/Standard 2022 is used to simulate the mechanical behavior of knit fabrics. The rest configuration of the knit geometry is described by a space curve parameterized along its arc length. A 2$\times$2 representative unit is defined and meshed using quadratic tetrahedral elements. Two static loading steps with nonlinear geometry are defined: The first uses thermal expansion to prevent any interference while allowing proper contact between the yarns. The second applies prescribed displacements in the two directions. This procedure is detailed in SI Section 7.

\section{Design of the sleeve}
\label{app5}
The target geometry is built from a 3D scan of the arm, while the initial geometry is created as a cylinder matching the wrist circumference. We generate structured quadrilateral meshes with identical numbers of elements for both geometries, establishing one-to-one mapping and estimation of resulting strains on each mesh. Initially, cotton yarn is assigned to all mesh elements; however, due to peak stresses exceeding cotton's yield stress, certain high-strain regions are selectively replaced with the nylon yarn in the rib pattern, the most compliant option available in the design space. Using our homogenized model, we predict stress from the optimized knit parameters—including stitch length, knit pattern, and yarn material—at the estimated strains. These parameters are optimized to minimize stress variance across the sleeve and refined to ensure contiguous parameter regions. Details of the optimization process are detailed in SI Section 9, along with fabrication and experimental validation methods. The final sleeve design is prepared with CREATE PLUS (Stoll) software and fabricated using a Stoll CMS 330 knitting machine.

 \bibliographystyle{elsarticle-num} 
 \bibliography{bib_2}






\end{document}